\documentclass[conference]{IEEEtran}
\IEEEoverridecommandlockouts
\usepackage{cite}
\usepackage{amsmath,amssymb,amsfonts}
\usepackage{algorithmic}
\usepackage{graphicx}
\usepackage{textcomp}
\usepackage{xcolor}
\usepackage[utf8]{inputenc}
\usepackage{float}
\usepackage{lipsum}  
\usepackage{adjustbox} 
\usepackage{subfig} 

\def\BibTeX{{\rm B\kern-.05em{\sc i\kern-.025em b}\kern-.08em
    T\kern-.1667em\lower.7ex\hbox{E}\kern-.125emX}}
\begin{document}

\title{Characterization of Covid-19 Dataset using Complex Networks and Image Processing}
\author{
\IEEEauthorblockN{Josimar Edinson Chire Saire}
\IEEEauthorblockA{Institute of Mathematics and\\
Computer Science (ICMC) \\
University of São Paulo (USP)\\
São Carlos, SP, Brazil\\
jecs89@usp.br}

\and

\IEEEauthorblockN{Esteban Wilfredo Vilca Zuñiga}
\IEEEauthorblockA{Dept. of Computing and Mathematics (FFCLRP)\\
University of São Paulo (USP)\\
Ribeirão Preto, Brasil\\
evilcazu@usp.br}
}

\maketitle

\begin{abstract}
This paper aims to explore the structure of pattern behind covid-19 dataset. The dataset includes medical images with positive and negative cases. A sample of 100 sample is chosen, 50 per each class. An histogram frequency is calculated to get features using statistical measurements, besides a feature extraction using Grey Level Co-Occurrence Matrix (GLCM). Using both features are build Complex Networks respectively to analyze the adjacency matrices and check the presence of patterns. Initial experiments introduces the evidence of hidden patterns in the dataset for each class, which are visible using Complex Networks representation. 

\end{abstract}


\section{Introduction}

Covid-19 is a breakthrough in human history. It is destroying powerful economies and collapsing emerge countries. During its early stage, the virus had a reproduction number of 4.22 in Germany and the Netherlands. Even if the developed countries reduced the impact of the virus, in countries with poverty and weak health systems, the virus is still a severe problem. By consequence, many efforts are focused on finding a vaccine, study the virus and find automatic tools to support prognosis of the illness.

Oriented on this direction, many groups are working with tomography, x-ray images to build a model using Artificial Intelligence techniques, i.e. Deep Learning \cite{Apostolopoulos2020} \cite{OZTURK2020103792}\cite{MAHMUD2020103869} At the same time a limitation in the first months was access to images related to covid-19 patients. Deep Learning algorithms are based in Artificial Neural Networks with many layers and where each layer or groups has an specific function but one limitation is the need of great quantity of images. For the previous reason, data augmentation is common to get artificial images with rotation, some noise.

One approach which meaningful feature extraction which represents internal patterns from one dataset is Complex Networks \cite{carneiro2018importanceconcept} \cite{seyed2019} \cite{chiresaire2020new} \cite{esteban2020bc}, previous experiments showed the strength of approaches using graph representation in comparison to classical Machine Learning algorithms. The presented results introduces the idea of good representation of the internal patterns. Besides, it is possible to affirm that this Complex Network representation can represent this patterns using a small number of samples in comparison with Deep Learning. 


In this paper, we propose a new technique based in Complex Networks to identify the virus in x-ray images using high-level algorithms to exploit the structure of the features from these images.


\section{Data and Methods}

\begin{figure}[hbpt]
\centerline{
\includegraphics [width=0.5\textwidth]{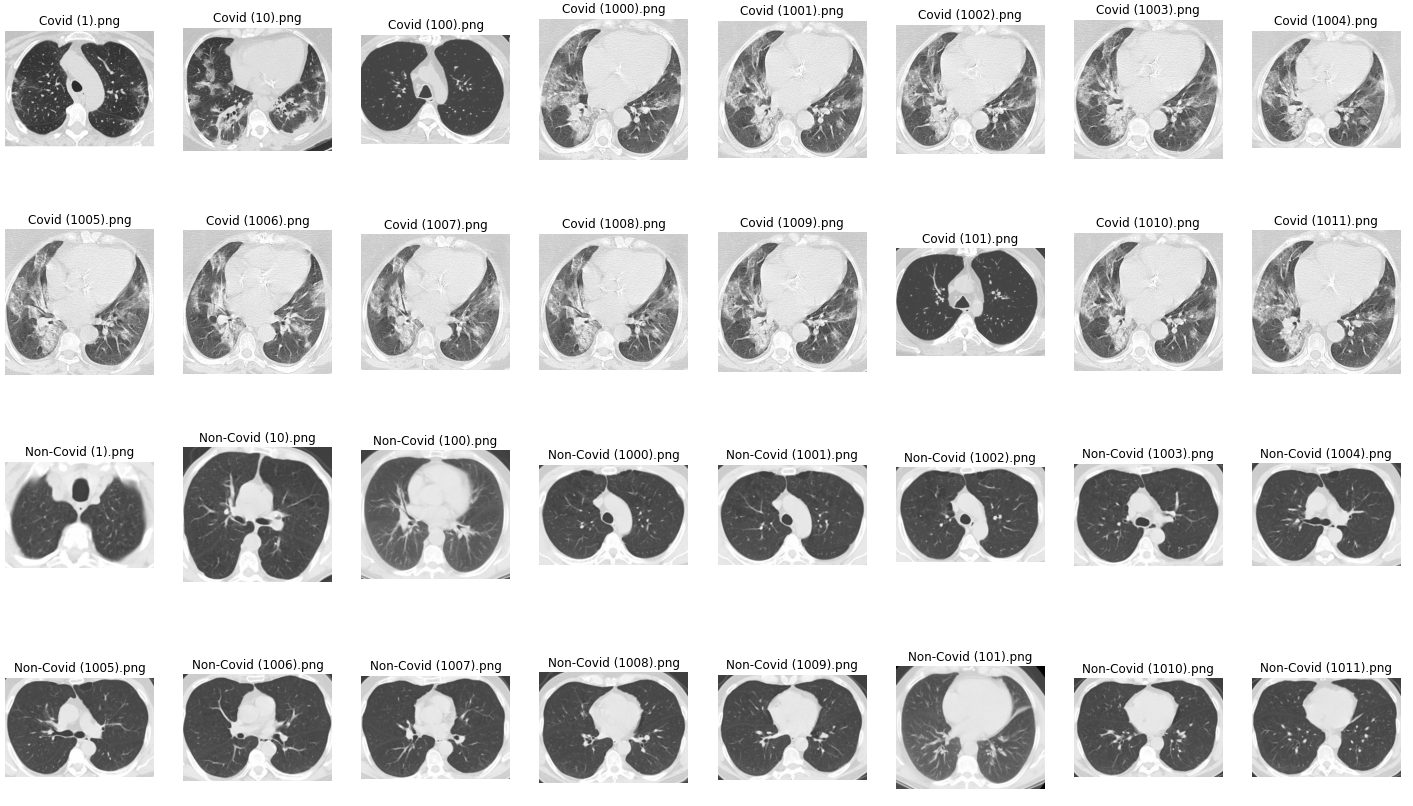}}
\caption{Sample of Dataset}
\label{fig:dataset}
\end{figure}

After a search using keywords, i.e. covid-19 tomography dataset. Many datasets were found but these are too big to download and process later, besides a variety of image formats are available, i.e. nii, dicom. One available dataset\footnote{https://www.kaggle.com/plameneduardo/sarscov2-ctscan-dataset/notebooks} is chosen because format png is ready for processing this images. The image Fig. \ref{fig:dataset} presents 16 samples of positive and negative cases respectively.

A dataset with 100 images is selected, positives and negative classes are balanced. By consequence is necessary to find a transformation from images to Complex Networks. A first proposal is using Frequency Histogram, because it can reduce dimensionality and represent the distribution of pixels. Previously, a transformation of color images is performed to get grayscale images. Later, a proposal using GLCM is done to get neighborhood features considering texture analysis.

\clearpage

\subsection{Frequency Histogram}

\begin{figure}[h]
\centerline{
\includegraphics [width=0.5\textwidth]{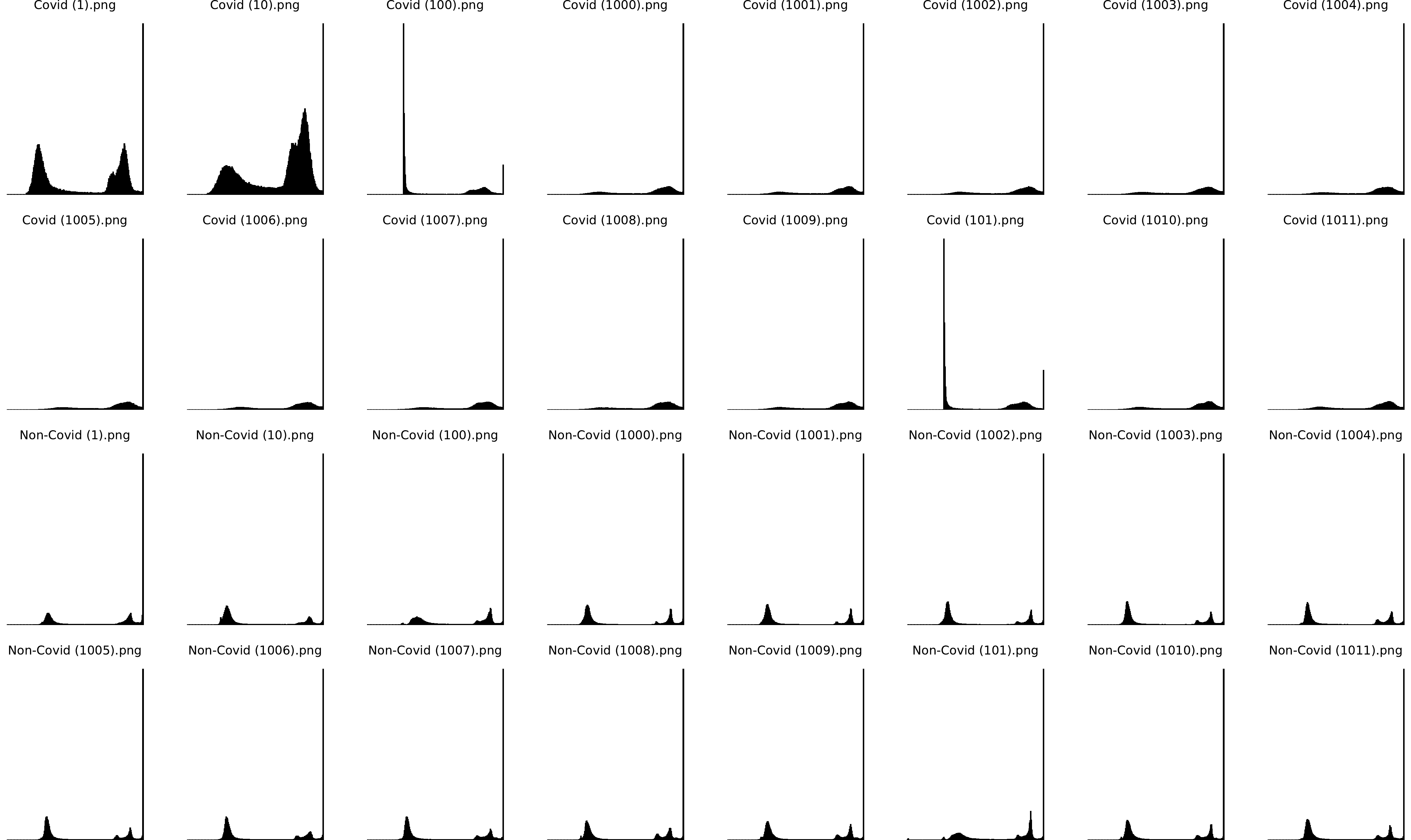}}
\caption{Characterization of dataset using Frequency Histogram}
\label{fig:histo}
\end{figure}

Histogram frequency were calculated to have a lower dimensionality representation and statistical features were calculated, median, mean, standard deviation, kurtosis and skew. This histogram is considering the three channels of classical RGB image representation. Figure Fig. \ref{fig:histo} represents the histogram frequency from the previous sample of images. Using this representation lets find a visual difference between positive and negative cases.


\begin{figure}[h]
\centerline{
\includegraphics [width=0.5\textwidth]{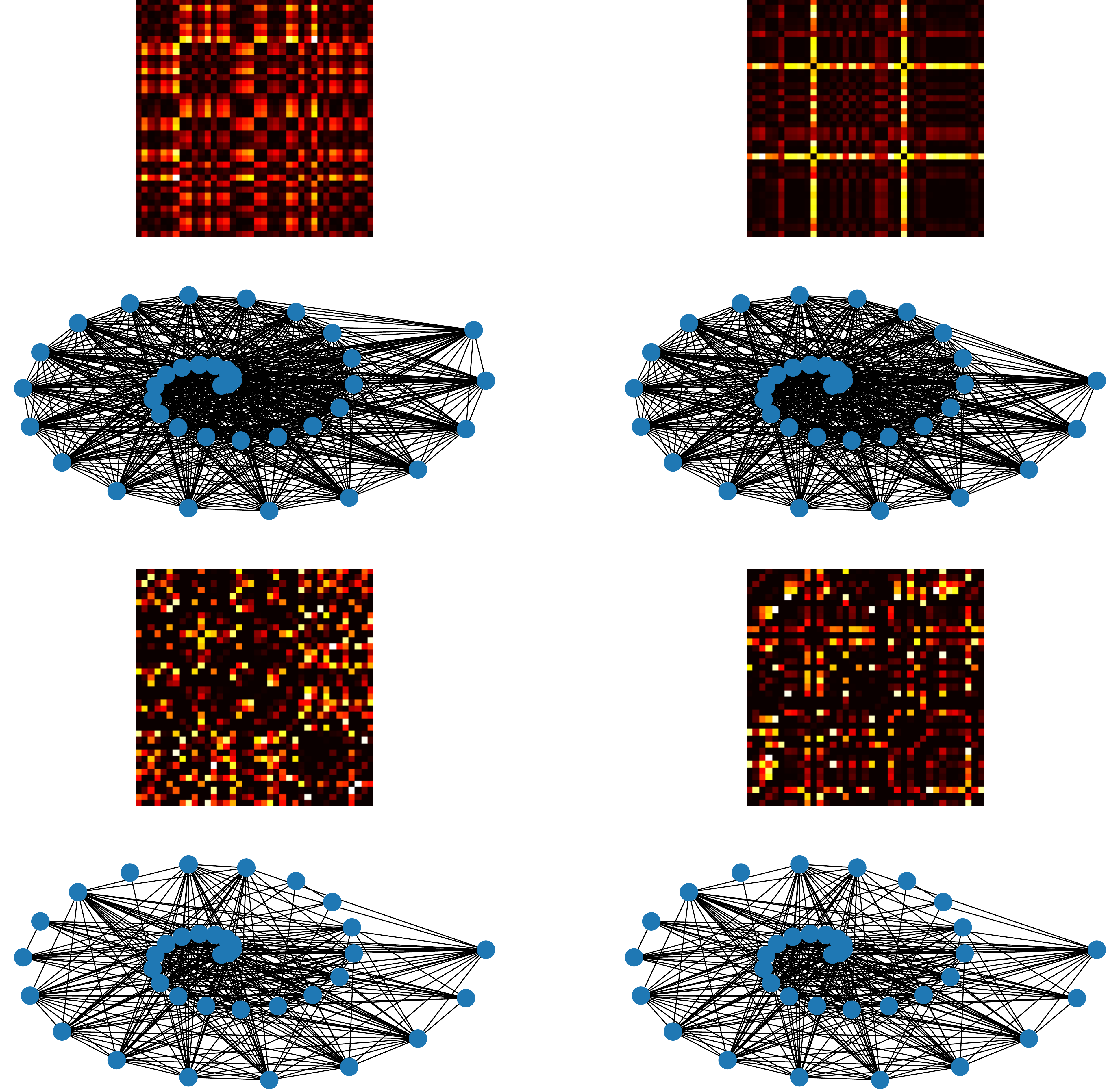}}
\caption{Characterization of dataset using Frequency Histogram}
\label{fig:cn}
\end{figure}

Then, the creation of graph is using euclidean distance between points of the same class, see Fig. \ref{fig:cn}. Left side represents positive class and right, negative ones. A filtering was performed using median of the distances, the bottom of the image introduces the results.

\subsection{Grey Level Co-Occurrence Matrix features}

Grey Level Co-Occurrence Matrix (GLCM) algorithm \cite{Muhammad} is a second order statistical method use for texture feature extraction. From this matrix, the next features are extracted:

\begin{itemize}
\item $contrast: \sum_{i,j}^{levels-1} P_{i,j}(i-j)^2$
\item $dissimilarity: \sum_{i,j}^{levels-1} P_{i,j}\|i-j\|$
\item $homogenity: \sum_{i,j}^{levels-1} \frac{P_{i,j}}{1+(i-j)^2}$
\item $ASM: \sum_{i,j}^{levels-1} P_{i,j}^2$
\item $energy: \sqrt(ASM)$
\item $correlation: \sum_{i,j}^{levels-1} P_{i,j}\frac{(i-\mu_i)(j-\mu_j)}{sqrt(\sigma_{i}^2 \sigma_{j}^2)}$
\end{itemize}

This features are considering 4 orientations: 0, 45, 90 and 135 degrees. A sample of the dataset is presented in Tab. \ref{table:dataset}. A transformation from RGB representation to grayscale is performed using this formula: 

\begin{equation}
    Image(i,j) = 0.3 * R + 0.59 * G + 0.11 * B, 
\end{equation}

where RGB are the red, gren, blue channels of the image.

\begin{table*}[!h]

\caption{GLCM Features}
\resizebox{0.80\textwidth}{!}{\begin{minipage}{\textwidth}
\begin{tabular}{|r|l|l|l|l|l|l|l|l|l|l|l|}
\hline
\textbf{}                   & \multicolumn{1}{c|}{\textbf{0}} & \multicolumn{1}{c|}{\textbf{1}} & \multicolumn{1}{c|}{\textbf{2}} & \multicolumn{1}{c|}{\textbf{3}} & \multicolumn{1}{c|}{\textbf{4}} & \multicolumn{1}{c|}{\textbf{...}} & \multicolumn{1}{c|}{\textbf{95}} & \multicolumn{1}{c|}{\textbf{96}} & \multicolumn{1}{c|}{\textbf{97}} & \multicolumn{1}{c|}{\textbf{98}} & \multicolumn{1}{c|}{\textbf{99}} \\ \hline
\textbf{dissimilarity\_0}   & \textbf{25.883713}              & 23.155504                       & 16.057523                       & 23.055059                       & 23.792782                       & ...                               & 14.738059                        & 18.991853                        & 15.172721                        & 14.429029                        & 13.812741                        \\ \hline
\textbf{dissimilarity\_45}  & \textbf{26.738676}              & 24.806191                       & 17.051322                       & 24.552878                       & 24.704867                       & ...                               & 16.286686                        & 18.880901                        & 16.447053                        & 14.804682                        & 13.781349                        \\ \hline
\textbf{dissimilarity\_90}  & \textbf{25.823644}              & 22.421290                       & 16.307116                       & 22.496803                       & 22.163044                       & ...                               & 14.585419                        & 17.292211                        & 14.963880                        & 13.419224                        & 11.963994                        \\ \hline
\textbf{dissimilarity\_135} & \textbf{27.982243}              & 23.578820                       & 17.093462                       & 23.276331                       & 22.817779                       & ...                               & 15.247283                        & 18.741751                        & 16.237618                        & 14.634247                        & 13.554673                        \\ \hline
\textbf{correlation\_0}     & \textbf{0.816670}               & 0.810789                        & 0.888780                        & 0.780967                        & 0.756747                        & ...                               & 0.905216                         & 0.883157                         & 0.903064                         & 0.906432                         & 0.911844                         \\ \hline
\textbf{correlation\_45}    & \textbf{0.816058}               & 0.790394                        & 0.876655                        & 0.758024                        & 0.747884                        & ...                               & 0.885158                         & 0.889271                         & 0.886499                         & 0.902797                         & 0.913434                         \\ \hline
\textbf{correlation\_90}    & \textbf{0.831263}               & 0.830631                        & 0.887386                        & 0.798635                        & 0.802751                        & ...                               & 0.908748                         & 0.910569                         & 0.906260                         & 0.920907                         & 0.936309                         \\ \hline
\textbf{correlation\_135}   & \textbf{0.800997}               & 0.809964                        & 0.875376                        & 0.781326                        & 0.783196                        & ...                               & 0.901909                         & 0.890089                         & 0.891492                         & 0.905173                         & 0.917023                         \\ \hline
\textbf{homogeneity\_0}     & \textbf{0.107910}               & 0.085270                        & 0.309934                        & 0.067542                        & 0.068566                        & ...                               & 0.190816                         & 0.092633                         & 0.171215                         & 0.202592                         & 0.205851                         \\ \hline
\textbf{homogeneity\_45}    & \textbf{0.090083}               & 0.080614                        & 0.306091                        & 0.062024                        & 0.064967                        & ...                               & 0.180754                         & 0.088947                         & 0.161639                         & 0.185911                         & 0.195601                         \\ \hline
\textbf{...}                & \textbf{...}                    & ...                             & ...                             & ...                             & ...                             & ...                               & ...                              & ...                              & ...                              & ...                              & ...                              \\ \hline
\textbf{contrast\_135}      & \textbf{2343.495992}            & 1477.278851                     & 1312.401532                     & 1317.666332                     & 1292.048722                     & ...                               & 1000.266048                      & 1017.557082                      & 1063.367497                      & 979.189887                       & 855.628087                       \\ \hline
\textbf{ASM\_0}             & \textbf{0.001265}               & 0.000418                        & 0.027868                        & 0.000213                        & 0.000224                        & ...                               & 0.001398                         & 0.000388                         & 0.001182                         & 0.001207                         & 0.001317                         \\ \hline
\textbf{ASM\_45}            & \textbf{0.000535}               & 0.000432                        & 0.027333                        & 0.000202                        & 0.000210                        & ...                               & 0.001347                         & 0.000364                         & 0.001119                         & 0.001154                         & 0.001288                         \\ \hline
\textbf{ASM\_90}            & \textbf{0.000520}               & 0.000431                        & 0.027970                        & 0.000212                        & 0.000228                        & ...                               & 0.001367                         & 0.000386                         & 0.001173                         & 0.001217                         & 0.001375                         \\ \hline
\textbf{ASM\_135}           & \textbf{0.000526}               & 0.000391                        & 0.027936                        & 0.000207                        & 0.000229                        & ...                               & 0.001356                         & 0.000392                         & 0.001120                         & 0.001156                         & 0.001278                         \\ \hline
\textbf{energy\_0}          & \textbf{0.035566}               & 0.020450                        & 0.166936                        & 0.014603                        & 0.014956                        & ...                               & 0.037393                         & 0.019696                         & 0.034384                         & 0.034747                         & 0.036287                         \\ \hline
\textbf{energy\_45}         & \textbf{0.023127}               & 0.020789                        & 0.165326                        & 0.014195                        & 0.014484                        & ...                               & 0.036701                         & 0.019084                         & 0.033447                         & 0.033966                         & 0.035886                         \\ \hline
\textbf{energy\_90}         & \textbf{0.022796}               & 0.020761                        & 0.167243                        & 0.014574                        & 0.015087                        & ...                               & 0.036970                         & 0.019646                         & 0.034255                         & 0.034892                         & 0.037086                         \\ \hline
\textbf{energy\_135}        & \textbf{0.022930}               & 0.019781                        & 0.167140                        & 0.014371                        & 0.015117                        & ...                               & 0.036819                         & 0.019808                         & 0.033463                         & 0.033995                         & 0.035746                         \\ \hline
\textbf{label}              & \textbf{1.0}                    & 1.0                             & 1.0                             & 1.0                             & 1.0                             & ...                               & 0.0                              & 0.0                              & 0.0                              & 0.0                              & 0.0                              \\ \hline
\end{tabular}
\end{minipage}}
\label{table:dataset}
\end{table*}

\begin{figure}[hbpt]
\centerline{
\includegraphics [width=0.5\textwidth]{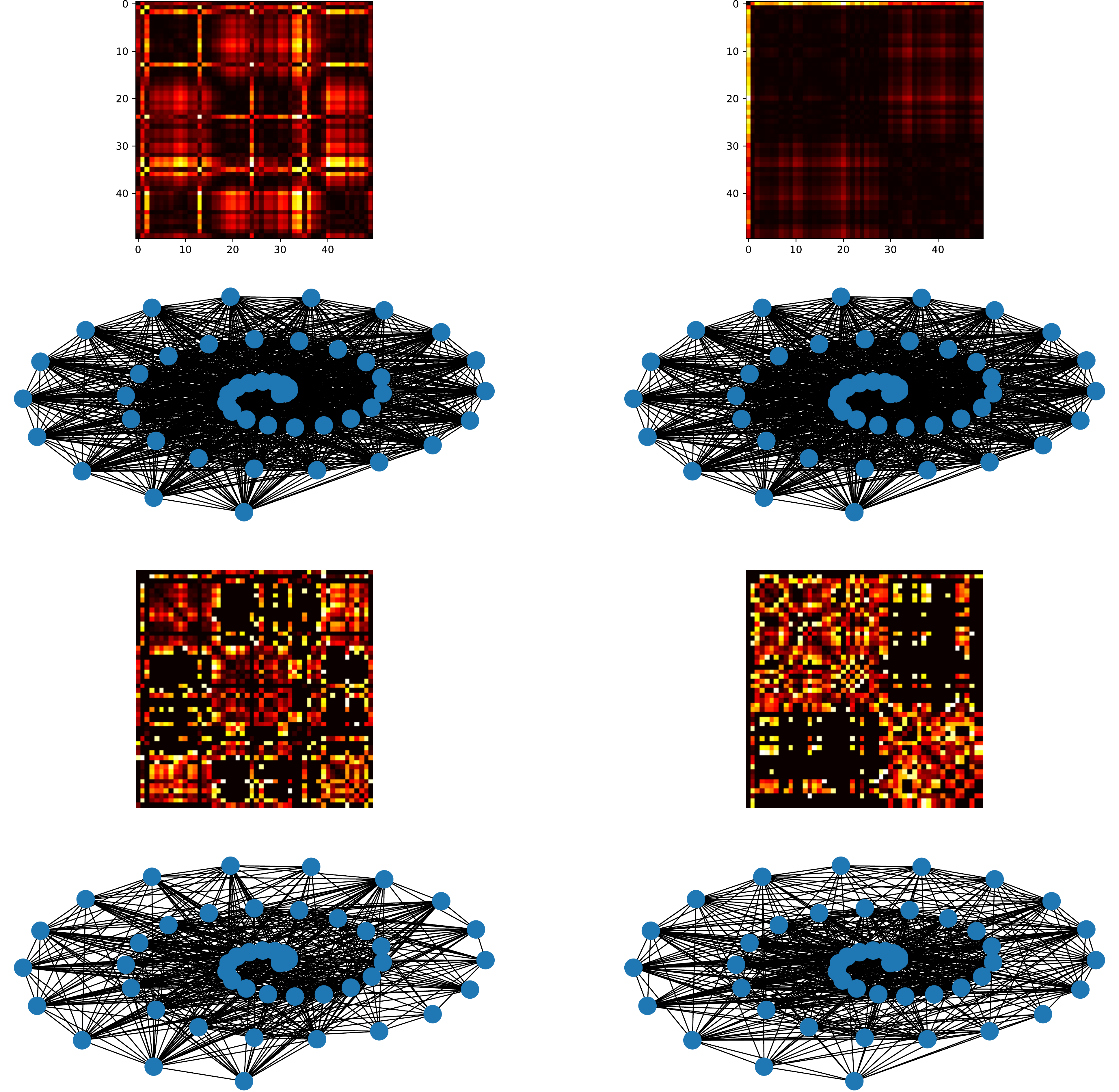}}
\caption{Characterization of dataset using GLCM}
\label{fig:glcm}
\end{figure}

Figure \ref{fig:glcm} presents the results for GLCM features. Column 1 is showing positive cases, and column 2, the negative ones.

Considering previous results, using Frequency Histogram and GLCM is possible to notice that Complex Networks building is possible using euclidean distance. Besides, the representation of Complex Network through adjacency matrices presents reticular patterns. This patterns are different, positive cases presents a distribution of further or higher distances between the nodes/elements than negative ones. By contrast, negatives samples presents only a few link with high distances. 

\section{Conclusions}

An approach to represent covid-19 tomography images using Complex Networks is feasible. The intensity of the links represented through adjacency matrices presents a strong difference between both classes. In spite of GLCM is a more elaborated technique to extract neighborhood pattern from the images, frequency histogram has a similar representation. Although, both processes are different to create the Complex Networks has a similar behaviour, this feature is presented in visualization of adjacency matrices. Besides, a comparison between Complex Networks approaches for High Level Classification will be presented.

\section{Future Work}

The authors are considering to include a higher number of samples for each classes to have higher diversity of images. Complex Networks representation can be leverage for High Level Classification tasks, then experiment on that way will be performed.

\section{Acknowledgments}

Authors wants to thank Research4tech, an Artificial Intelligence(AI) community of Latin American Researcher with the aim of promoting AI, build Science communities to catapult and enforce development of Latin American countries supported on Science and Technology, integrating academic community, technology groups/communities, government and society.

\bibliographystyle{IEEEtran}

\bibliography{biblio.bib}

\vspace{12pt}
\end{document}